# Research on the photoacoustic spectrum analysis using k-Wave

Xianlin Song [a, #, *], Jianshuang Wei [b, c, #], Qi Jiang [a], Lingfang Song [d]
[a] School of Information Engineering, Nanchang University, Nanchang 330031, China;
[b] Britton Chance Center for Biomedical Photonics, Wuhan National Laboratory for Optoelectronics-Huazhong University of Science and Technology, Wuhan 430074, China;
[c] Moe Key Laboratory of Biomedical Photonics of Ministry of Education, Department of Biomedical Engineering, Huazhong University of Science and Technology, Wuhan 430074, China;
[d] Nanchang Normal University, Nanchang 330031, China;
[#] equally contributed to this work

## ABSTRACT

Photoacoustic imaging is a new non-destructive medical imaging technology based on photoacoustic effect. It can reflect the difference of light absorption energy by detecting photoacoustic signal. At present, the analysis methods of photoacoustic signals in biological tissues can be divided into two categories, namely, time-domain analysis of signals and frequency-domain analysis of signals. In time domain analysis, the envelope of the received photoacoustic signal is usually used to reconstruct the image. However, due to the influence of various external factors, the time domain signal cannot accurately reflect the characteristics of the absorber itself. Here, photoacoustic spectrum analysis was performed by using k-Wave to obtains the relationship between the structure, size, density of the absorber and the photoacoustic spectrum. Firstly, the relationship between the size of absorber and the photoacoustic spectrum is studied, and the slope and intercept are used to analyze the spectrum. Conversely, the relationship was used to predict the size of the absorber Finally, we used this relationship to predict the size of blood vessels.

## 1. INTRODUCTION

Photoacoustic imaging is a promising technique that combines optical contrast with ultrasonic detection to map the distribution of the absorbing pigments in biological tissues [1-4]. It has been widely used in biological researches, such as structural imaging of vasculature [5], brain structural and functional imaging [6], and tumor detection [7]. Considering the lateral resolution of photoacoustic microscopy (PAM), it can be classified into two categories: optical-resolution (OR-) and acoustic-resolution (AR-) PAM [8, 9]. In AR-PAM, the spatial resolution is determined by the acoustic focus, since the laser light is weekly or even not focused on the sample. Conversely, in the OR-PAM, the laser light is tightly focused into the sample to achieve sharp excitation.

At present, the analysis methods of photoacoustic signals in biological tissues can be mainly divided into two categories: signal time domain analysis and signal frequency domain analysis. The researchers hope that by analyzing the signals in the time and frequency domains, they can obtain information about various aspects of the organization.

In time domain, the envelope of photoacoustic signal reflects the distribution of light absorbers in tissue. Intuitively, the wider the envelope and the larger the amplitude of the signal, the larger the size of the absorber, the stronger the light absorption capacity. The reconstructed image of the time domain signal is determined by the relative intensity of the absorber and the amplitude of the photoacoustic signal. This is very similar to ultrasound imaging, except that the grayscale image of the ultrasound reconstruction represents the acoustic impedance information in the tissue. In recent years, many researches focus on the time domain analysis of photoacoustic signals, such as image reconstruction. At pre-sent, many algorithms have been proposed at home and abroad. One of the earliest algorithms was the Radon transform. Typically, the signal is measured by cyclic scanning. When the object to be measured is in the scanning area and the size of the object to be measured is small relative to the scanning orbit, the reconstruction formula can be approximate to the linear Radon transform method in CT imaging. Subsequently, Kruger group, Liu group and Xu group respectively studied the algorithm [10, 11].In general, however, the reconstructed absorber needs to be near the center of the scanned orbit. In 2002, Xu et al. developed an algorithm for backprojection in the time domain and obtained an analytical expression in the case of spherical scanning [12]. Based on this algorithm, nondestructive imaging of the structure and function of the mouse brain was performed in 2003.Then, on this basis, they also developed a more general reconstruction algorithm, which has been widely used [13, 14].Of course, there are also many other algorithms, such as image reconstruction by solving matrix equations [15].The starting point of the above method is from the time domain analysis of the signal.

* songxianlin@ncu.edu.cn; phone 0086-791-83969675; http://ies.ncu.edu.cn

In the frequency domain, the larger size of the optical absorber corresponds to a narrower spectrum bandwidth, while the smaller size of the optical absorber corresponds to a wider spectrum bandwidth. The stronger the light absorption effect of a light absorber is, the higher the total energy of its corresponding power spectral density will be. At present, the study of photoacoustic signal in frequency domain can be divided into two aspects. One is to analyze the problem from the perspective of application. Silverman et al. took out the eye tissue and conducted in vitro photoacoustic experiments, and analyzed the high frequency part of the photoacoustic signal of the tissue. The results showed that the medium frequency band and slope of the signal power spectrum would change in the colored iris [16].Kumon et al. conducted frequency domain analysis on prostate cancer in mice and found that there were significant differences in frequency domain signal characteristics (such as intermediate frequency, slope, intercept, etc.) between cancerous and non-cancerous regions [17].On the other hand, it is a theoretical analysis. Kolio, whelan et al. conducted a large number of studies on the photoacoustic spectrum of blood cells [19].They used simulations to simulate what would happen if randomly distributed blood cells were clustered and unclustered. It turns out that when blood cells are in the aggregation state, it corresponds to a narrower spectrum of signals. In general, so far, there are few studies on photoacoustic signals in the frequency domain at home and abroad, and most of them are only qualitative analysis problems, which are far from enough compared with the studies on time domain signals.

As mentioned earlier, for time domain analysis, the envelope of the received photoacoustic signal is usually used to reconstruct the image. However, the time-domain signal is affected by various external factors, such as the excitation of the light source, the response of the sensor, and the transfer function of the measurement system. These factors are external factors, which leads to the reconstruction image can only qualitatively reflect the relative intensity of the absorber, but not quantitatively reflect the properties of the absorber itself. That is to say, for the same sample, the results obtained by different instruments or even different operators of the same instrument cannot be compared quantitatively. So we want to have a method of measurement that is independent of the instrument and the operator. If this can be done, it will provide great convenience for the diagnosis of the disease.

In this paper, we developed a photoacoustic spectrum analysis platform using k-Wave to obtains the relationship between the structure, size, density of the absorber and the photoacoustic spectrum. And, conversely, we used the relationship to predict the structural characteristics of the absorber.

## 2. METHOD

### 2.1 Configuration Environment

The k-Wave simulation toolbox can analyze photoacoustic signals in the time domain [16]. We use k-Wave: MATLAB toolbox for the simulation of Bessel-beam photoacoustic microscopy. The simulated environment was created in three dimension with 200 x 200 x 200 voxels (each voxel size is 0.5 μm), as shown in Figure 1, and contains a perfectly matched boundary layer (PML) to satisfy the boundary conditions for the forward process. The surrounding medium is water with a sound velocity of 1.5 km/s and a density of 1000 kg/m3. All simulations assume that an acoustically homogeneous medium was considered with no absorption or dispersion of sound.

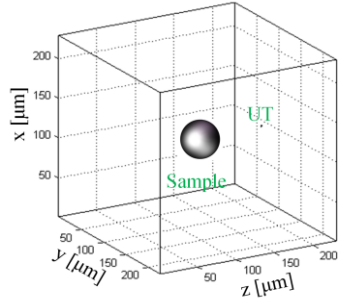

Figure 1. A phase diagram that produces double annular beams. The prism phase in the double annular slits deflects the beam from the optical axis.

In our simulation, the parameters of the simulated system were chosen to match those of the practical setup as closely as possible, the center frequency of the ultrasonic transducer is set to 50 MHz and the bandwidth is 80%. The three-dimensional imaging data can be obtained by carried out two-dimensional raster scan.

**2.2 Deconvolution eliminates ultrasonic probe bandwidth limitations**

In order to ensure that the obtained signal is the original signal, the photoacoustic signal is deconvolution processed. Deconvolution is a basic problem in signal processing, which is widely used in channel equalization, image restoration, speech recognition, seismic nondestructive flaw detection and other fields.In general, the purpose of deconvolution is to find a solution to the convolution equation of a form:

$$f * h = g \qquad (1)$$

Where $*$ represents the convolution operation. $h$ is the system response function, $f$ is some signals we want to recover, but before we record, it has been convolved with some other signals, if we know $g$ , then we can perform deterministic deconvolution.

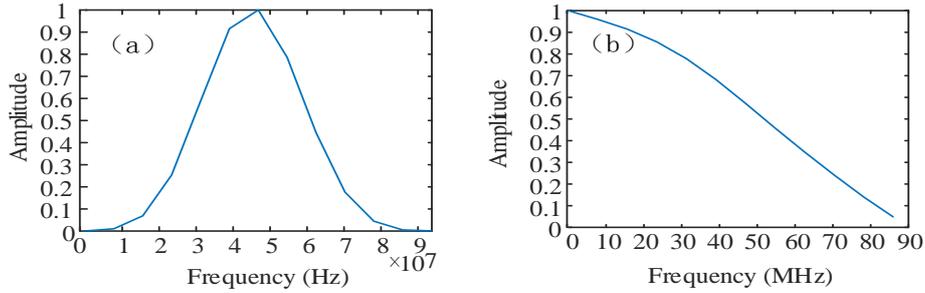

Figure 2. Signal spectrum before (a) and after (b) deconvolution.

Therefore, in the actual photoacoustic image, the degradation of photoacoustic images can be divided into two reasons. One is due to the existence of noise in the observation process, which leads to the change of signal-to-noise ratio, so that the quality of the reconstructed image becomes low. Another reason is that the signal used for image reconstruction is obtained by the convolution of the impulse response of the system and the ultrasonic signal generated by the absorber Therefore, it is necessary to eliminate the influence of noise and system impulse response in the process of photoacoustic imaging. Figure 2 shows the signal spectrum diagram of a ball with a radius of 10 μm before (Figure 2 (a)) and after (Figure 2 (b)) deconvolution, so that the changes before and after deconvolution can be clearly compared.

# 3. RESULTS

## 3.1 The relationship between photoacoustic signal spectrum and size.

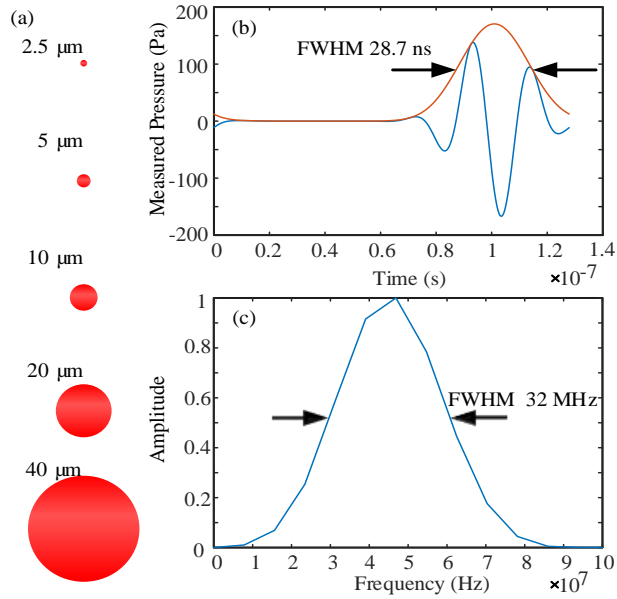

Figure 3. The relationship between photoacoustic signal spectrum and size.

Figure 3 is an experiment on the relationship between photoacoustic signal spectrum and size. A total of five small ball were taken in the simulation. They are small balls with radii of 2.5 μm, 5 μm, 10 μm, 20 μm and 40 μm, as shown in Figure 3 (a). Figure 3 (b) is photoacoustic signal of the ball with a radius of 10 μm. The full width at half maximum (FWHM) of the signal envelope (Hilbert transform) is 28.7 ns. Figure 3 (c) is the photoacoustic spectrum corresponding to the photoacoustic signal, which reflects the bandwidth (pass frequency range) of the system. It can be seen that the pass frequency range of the system is 0-100 MHz and the bandwidth is 32 MHz.

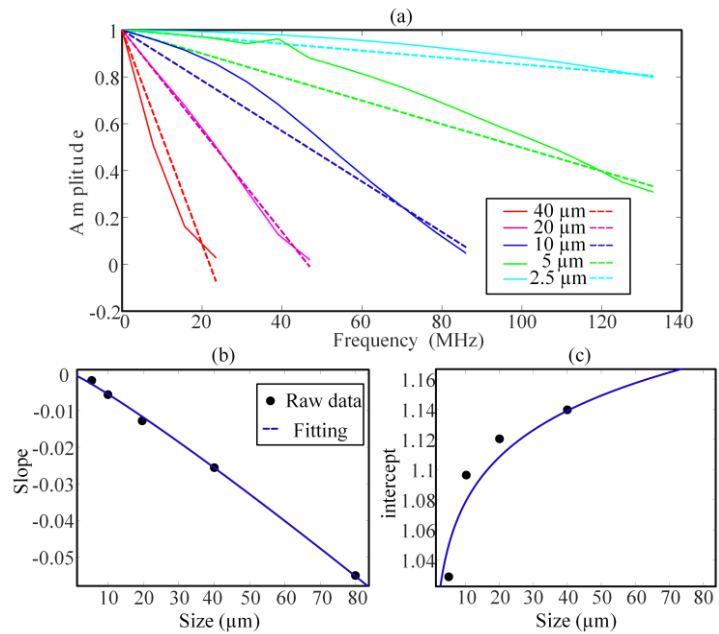

Figure 4. Spectrum diagram.

After recording the simulated photoacoustic spectrum curve of the five small balls, the curve is fitted, as shown in Figure 4(a). Draw a curve with the size as the abscissa and the slope and intercept as the ordinate to obtain the relationship between the frequency spectrum and the size, as shown in Figure 4(b) and Figure 4(c). It can be concluded that:

(1) The larger the size of the absorber, the more concentrated the signal spectrum is in the low-frequency region, the less high-frequency information, and the smaller the slope of the photoacoustic signal spectrum curve;
(2) The intercept and slope characteristics are opposite, as the size of the photoacoustic absorber increases, the intercept of the spectrum curve of the photoacoustic signal also increases.

Finally, 10 small balls were taken around 5 μm, 10 μm, and 20 μm, namely 1-10 μm, 6-15 μm, and 16-25 μm. Take the slope and intercept of the spectrum curve of the photoacoustic signal of these 30 small-size balls, divide them into three groups of data, each group of data includes 10 points, and use the k-means clustering algorithm for analysis. Obtain the slope and intercept of the three center points (the five-pointed star in the figure), as shown in Figure 5. And the data of these three center points (slope and intercept) are substituted into the relationship between the size and the frequency spectrum obtained. The predicted sizes are (16.075 μm, 16.139 μm), (10.02 μm, 10.114 μm) and (4.927 μm, 5.057 μm). The predicted size is basically consistent with the actual size, which also verifies the reliability of the conclusion.

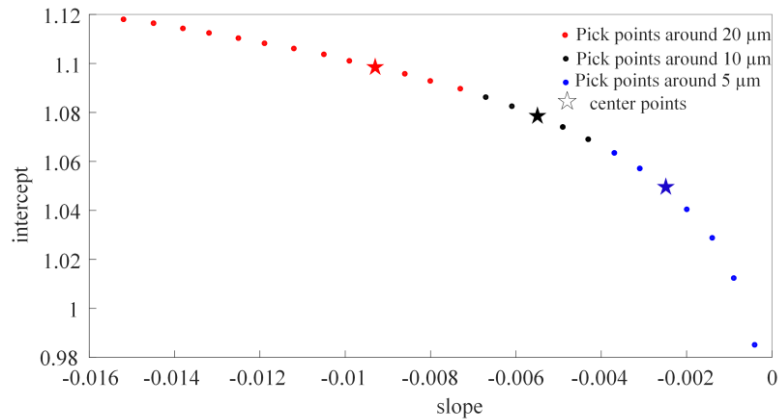

Figure 5. Cluster analysis graph.

## 4. CONCLUSION

In summary, we carried out photoacoustic spectrum analysis using k-Wave to obtains the relationship between the structure, size, density of the absorber and the photoacoustic spectrum. Firstly, the relationship between the size of absorber and the photoacoustic spectrum is studied, and the slope and intercept are used to analyze the spectrum. Conversely, the relationship was used to predict the size of the absorber. And the results show that the photoacoustic spectrum can effectively predict the size of the absorber. Finally, we used this relationship to predict the size of blood vessels, and the results demonstrate the effectiveness of our method.